# Anomalous Hall effect in an amorphous antiferromagnet with inverted hysteresis

Xiangning Du[#], Yuxiang Zhu[#], Na Chen[*]

School of Materials Science and Engineering, Tsinghua University, Beijing, China

# These authors contribute equally.
*Corresponding author: chennadm@mail.tsinghua.edu.cn

## Abstract

Stemming from antiferromagnetic coupling, exchange bias allows inverted hysteresis in a magnetic system. Such room temperature magnetic reversal has yet to be observed in an amorphous antiferromagnet. Furthermore, the impact of this exchange bias effect on its magnetoelectric transport behavior remains a mystery. Here we discovered a zero-field magnetization switching effect in an exchange-biased amorphous antiferromagnet with inverted magnetic hysteresis. This zero-field magnetic reversal was further evidenced by its inverted large anomalous Hall effect. Notably, this collective spin flipping at zero field can occur at room temperature or above room temperature, which may be associated with quantum interference effect due to thermal fluctuation enhanced disorder. Our experimental results offer a way to design room-temperature exchange-biased amorphous antiferromagnets with zero-field multi magnetic-states and large anomalous Hall effect, holding potential for low-power and high-density memory applications.

With further miniaturization of spintronic/electronic devices, the search for advanced magnetic materials with unique electronic and magnetic properties has evolved into an important field in materials science.[1-6] The exchange interaction among electrons, spins, and orbitals in these magnetic materials governs their overall magnetic, electronic, and transport properties. In particular, the coupling between ferromagnetic and antiferromagnetic layers introduces an additional anisotropy, which triggers exchange bias to stabilize specific magnetic configurations with collective spin reversal.[5] So far, the exchange bias effect has been observed in superparamagnets, ferrimagnets, and multilayer thin film interfaces. Despite the presence of ferromagnetic-antiferromagnetic interactions, the inverted hysteresis effect induced by exchange bias in antiferromagnets has yet to be discovered.

On the other hand, certain antiferromagnets with canted spins exhibit either zero or negligible magnetization, yet they behave like weak ferromagnets by displaying a large anomalous Hall effect and ultrafast magnon dynamics.[11-14] These results indicate that exchange bias could exist in antiferromagnets with specific magnetic configurations. Given the imperative scientific and technological importance, the design and development of exchange-biased antiferromagnetic systems hold promise for discovering exotic magnetoelectric transport properties. In the present study, an exchange-biased amorphous antiferromagnetic hybrid was developed, showing inverted hysteresis and anomalous Hall effect at room temperature. The zero-field magnetization switching effect at room temperature or above indicates a self-driven collective spin reversal possibly associated with a disorder-mediated quantum interference effect.

The magnetoelectric transport behavior of this amorphous system shows a magnetic transition from a low-temperature ferromagnetic state to a high-temperature antiferromagnetic state (**Figure 1**). Below 200 K, a large anomalous Hall effect is obtained, which is typical for a ferromagnet (**Fig. 1a**). Strikingly, an inverted anomalous Hall effect is detected at 200 K or above, which manifests antiferromagnetic behavior as observed in previous reports.[11-14] That is, this amorphous thin film acts as a room-temperature antiferromagnet, which transforms into a ferromagnet at temperatures below 200 K. Surprisingly, an irreversible anomalous Hall effect and zero-field magnetization-switching behavior have been observed at a higher temperature of 350 K.

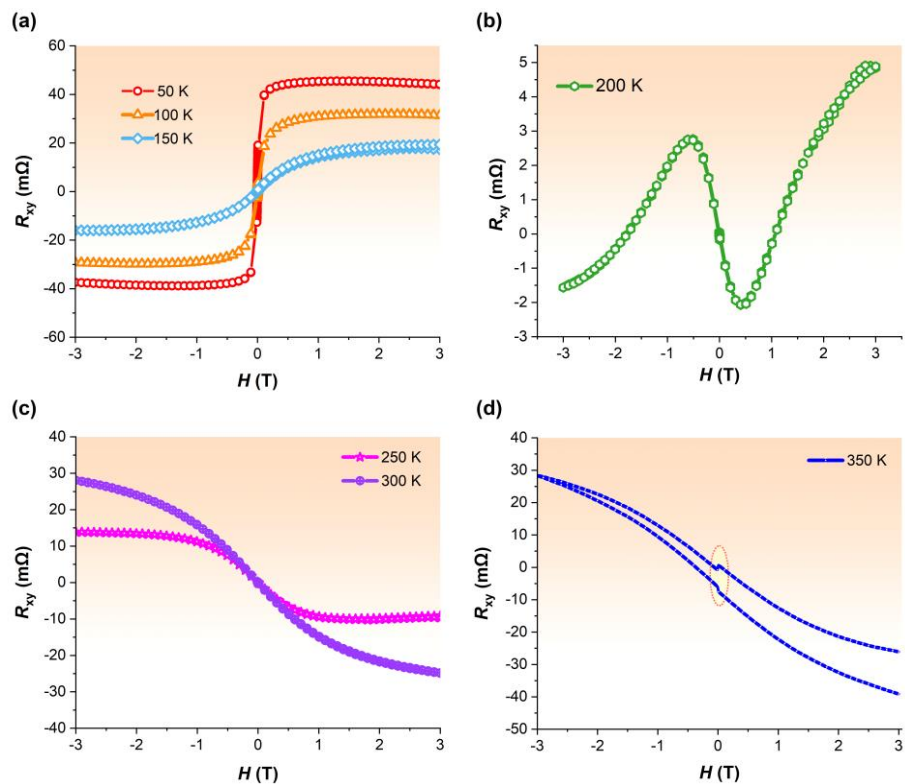

Fig. 1. Anomalous Hall effects of the room-temperatature amorphous antiferromagnet.

As depicted in **Figure 2a-c**, the magnetic hysteresis loops confirm the presence of a magnetic transition from a low-temperature ferromagnetic state to a high-temperature exchange-biased state. At temperatures of 250 K or higher, the magnetization hysteresis loop is inverted. Especially, a zero-field magnetization switching effect is observed at 350 K, leading to multiple magnetization states (**Fig.**

**2d**). This result agrees with the switching behavior at zero field in its anomalous Hall effect curve, as indicated in **Fig. 1d**. The existence of exchange bias gives rise to both parity-time symmetry breaking and leads to a collective spin flipping to decrease the overall ferromagnetism, which should be an energy-driven process.

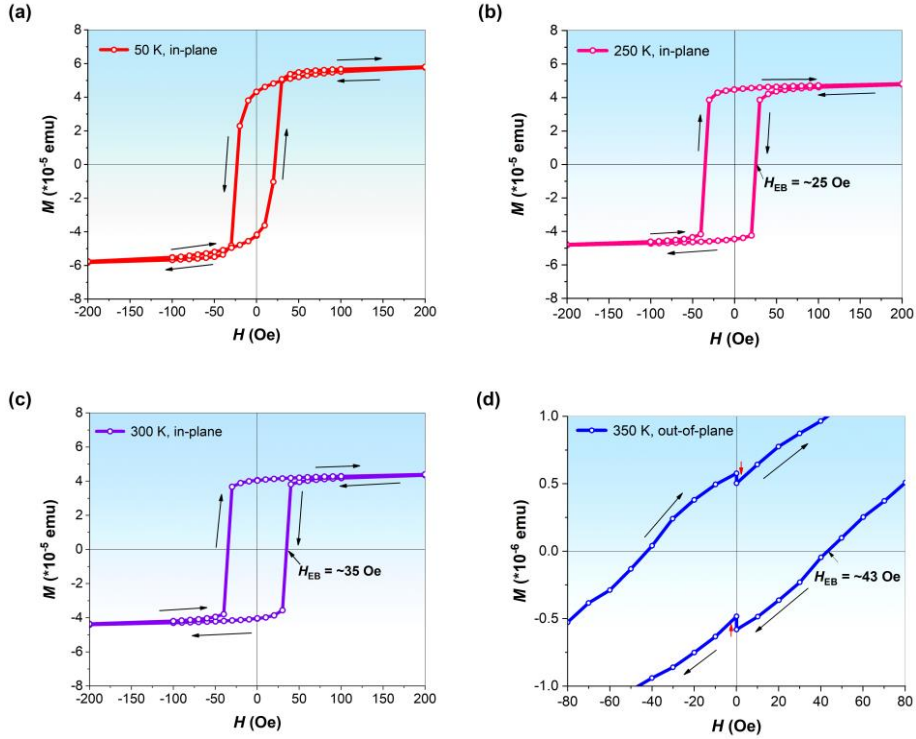

Fig. 2 Magnetization versus magnetic field (*M-H*) curves of the room-temperature amorphous antiferromagnet. (a) – (c) In-plane *M-H* curves showing inverted hysteresis at temperatures of 250 K and higher. (d) Out-of-plane *M-H* cureve at 350 K showing zero-field magnetization switching effect in consistent with that observed in the anomalous Hall effect measured at 350 K (Fig. 1d).

The transition from a low-temperature ferromagnetic state to a high-temperature antiferromagnetic state is accompanied by a change in its electrical behavior, as depicted in **Figure 3**. The resistance (*R*) versus temperature (*T*) curve exhibits nonlinear behavior at low temperatures and linear behavior at high temperatures (**Fig. 3a**). The derivative curve (d*R*/d*T*) shows a crossover at 250 K, as illustrated in **Fig. 3b**, which corresponds to the transition between these magnetic states.

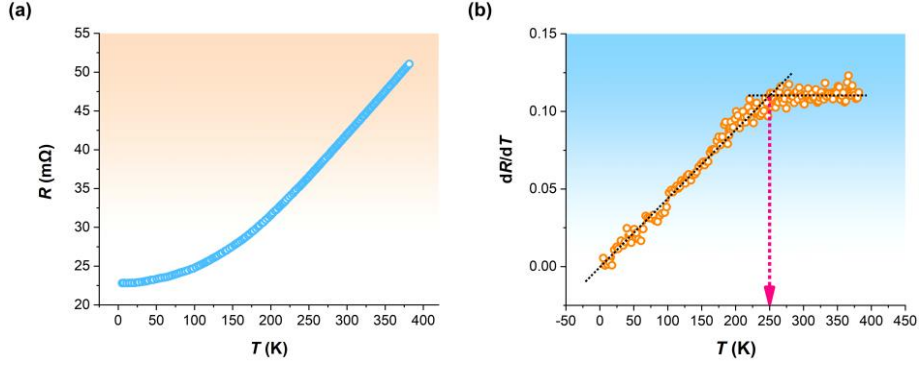

Fig. 3 (a) Resistance versus temperature (*R-T*) curve of the room-temperature amorphous antiferromagnet. (b) The derivative curve of the *R-T* curve, showing a crossover point of 250 K.

An exchange-biased system is featured by a ferromagnetic-antiferromagnetic interface exchange coupling[10]. A two-phase model has been used to probe the interplay between different magnetic states[6], which can be further extended to antiferromagnetic systems with canted spins. Assuming an antiferromagnetic hybrid system consisting of a ferromagnetic component (A) and an antiferromagnetic matrix component (B), the total free energy is given by the following equation.

$$E = -M_A H \cos\theta_A - M_B H \cos\theta_B - J\cos(\theta_B - \theta_A) + K_A \sin^2(\theta_A - \phi_A) + K_B \sin^2(\theta_B - \phi_B) \quad (1)$$

The first two terms represent the Zeeman energies of the two magnetic components, the third term denotes the exchange interaction energy induced by the interface magnetic coupling, and the last two terms correspond to the anisotropy energies of the two components. To minimize energy, the moment of the ferromagnetic component eventually rotates to align in parallel with the external field. Given a strong ferromagnetic-antiferromagnetic interaction, the net magnetization of the antiferromagnetic component could align antiparallel to the field, thereby reducing the energy. If the exchange interaction is sufficiently weak, the moments of the two components can eventually rotate along the field. In other words, the magnetism of such an exchange-biased system is tunable by adjusting the exchange strength *J*. At zero magnetic field, the total free energy $E$ is mainly determined by the exchange interaction energy induced by the interface antiferromagnetic coupling. Fundamentally, this antiferromagnetic coupling results in a decrease in the free energy, thus favoring zero-field magnetization switching through the rapid conversion of collective spins. These results suggest that manipulating $J$ in terms of the exchange bias strength ($H_{EB}$) in amorphous magnets offers an alternative way to create exchange-biased amorphous systems with tunable magnetic properties such as saturation magnetization ($M_s$).

As a result, an abnormal anisotropy effect in $M_s$ has been observed in this amorphous antiferromagnetic system as shown in **Figure 4**. $M_s$ of a magnetic system is defined as the maximum magnetization achieved when all of its local magnetic moments are aligned with an applied magnetic field. Regarded as an intrinsic parameter, its $M_s$ is supposed to be independent on the external magnetization directions. However, the relatively strong exchange strength $J$ stabilizes the antiferromagnetic configuration in specific magnetization direction, thereby leading to $J$-dependent anisotropic $M_s$ as shown in **Fig. 4a** ($\Delta M_s$ is defined as $M_{s,\perp} - M_{s,//}$, $M_{s,\perp}$ is the out-of-plane $M_s$, $M_{s,//}$ is the in-plane $M_s$). We normalize $\Delta M_s$ in different amorphous antiferromagnets by defining a reduced $\Delta M_s$ in terms of $\delta = \frac{\Delta M_s}{M_{s,//}}$. As shown in **Fig. 4b**, $\delta$ decreases with $H_{EB}$, confirming the crucial influence of exchange bias on the observed abnormal magnetic and magnetoelectrical transport properties in these exchange-biased amorphous antiferromagnets. Moreover, these abnormal behaviors can be controlled by adjusting $J$ in terms of $H_{EB}$.

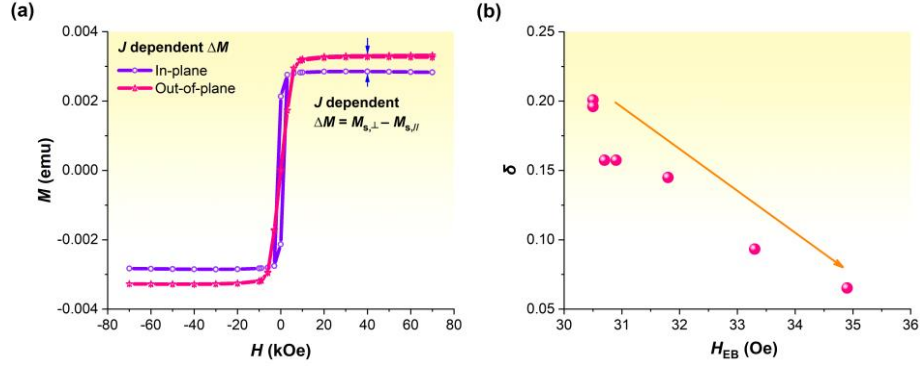

Fig. 4 (a) Anisotropic saturation magnetization observed in *M-H* curves of a room-temperature amorphous antiferromagnet. (b) Correlating room-temperature $\delta$ with $H_{EB}$ representing the exchange interaction strength.

A new magnetization switching effect has been observed at zero magnetic field in an exchange-biased amorphous antiferromagnetic system exhibiting inverted hysteresis and a large anomalous Hall effect. Remarkably, this collective spin flipping at zero field occurs even at room temperature and above, which may be linked to a quantum interference effect stemming from thermally enhanced disorder. The ability to switch multiple magnetization states at zero field through rapid collective spin conversion holds significant potential for the development of low-power, high-density memory technologies. Our findings suggest a method for producing new exchange-biased amorphous antiferromagnets characterized by a large anomalous Hall effect and inverted hysteresis.

**Acknowledgements**

This work was supported by the National Key Research and Development Program of China (2022YFA1402603). Support from Key Laboratory for Advanced Materials Processing Technology (MOE), Center for Testing and Analysing of